# MODELING OF THE CRYSTAL FIELD, MAGNETOELASTIC INTERACTIONS AND RANDOM LATTICE DEFORMATIONS IN $Pr_2Zr_2O_7$

V.V. Klekovkina, N.M. Abishev
Kazan Federal University
e-mail: vera.klekovkina@gmail.com

We present the results of simulations of the spectral and thermodynamic properties of frustrated $Pr_2Zr_2O_7$ crystals using the distribution function of random strains induced by point defects in an elastically anisotropic continuum.

## Introduction

Rare-earth oxides with the general formula $RE_2M_2O_7$ (where RE — rare-earth (RE) ion, M — tetravalent metal ion) are isostructural to pyrochlore mineral. Pyrochlore has a cubic face-centered lattice (space group Fd3m). In pyrochlores, the RE ion sublattice forms a network of vertex-connected tetrahedra. Four RE ions located at the vertices of one such tetrahedron are crystallographically equivalent, but magnetically nonequivalent. Such compounds are geometrically frustrated magnets and show a wide variety of magnetic behavior at low temperatures, thus attracting the attention of the scientific community. The magnetic properties of these compounds (in case where the metal ion is nonmagnetic) are determined by magnetic exchange and dipole interactions between RE ions and the interaction of RE ions with the crystal field (CF). RE ions are coordinated by eight oxygen ions located at the vertices of the distorted cube.

The main $^3H_4$ multiplet of the $Pr^{3+}$ ion in the trigonal CF of a perfect crystal with the pyrochlore structure splits into 3 $\Gamma_3^+$ doublets and 3 $2\Gamma_1^+ + \Gamma_2^+$ singlets (corresponding irreducible representations of the $D_{3d}$ point symmetry group in the position of praseodymium ions are indicated). In the CF of praseodymium zirconate, as was shown in [1], the ground electronic state of the $Pr^{3+}$ ion is a non-Kramers doublet. The first excited state (singlet) has an energy of the order of 110 K [1]. Modeling of the magnetic properties of praseodymium zirconate at low temperatures is usually performed using the effective spin Hamiltonian acting in the state space of the ground doublet.

In the experimental temperature dependences of the heat capacity $Pr_2Zr_2O_7$ [1–4], there is a Schottky anomaly near the temperature 2 K, which indicates splitting of the main doublet of the $Pr^{3+}$ ion. The presence of splitting of the main doublet and the random nature of its magnitude are indicated by the intensity maxima of inelastic neutron scattering observed at low temperatures in the low-energy part of the spectrum [4]. The nature of the interactions leading to this splitting has not been explained to date. The modeling of the properties of $Pr_2Zr_2O_7$ taking into account the splitting of the main doublet was carried out in [5–7]. The splitting value was assumed to be constant [5] or random with a one-dimensional distribution function [6, 7]. Nevertheless, both of these assumptions do not have any physical justification.

A characteristic feature of crystals with the pyrochlore structure is their nonstoichiometric [8], which arises when a part of the metal ions $M^{4+}$ are replaced by RE ions (in this case, additional vacancies are formed in the anionic oxygen sublattice) and when a part of the $RE^{3+}$ ions is replaced by $M^{4+}$ metal ions. In general, the formula of crystals with the pyrochlore structure (including crystals that are synthesized initially as stoichiometric ones) can be written in the form $RE_{2+x}M_{2-x}O_{7-x/2}$ (model "stuffed pyrochlore").

In the [9] heat capacity and magnetic susceptibility of $Pr_2Zr_2O_7$ were measured for a series of samples obtained under different synthesis conditions (growth rate, annealing time). The observed dependence of the measurement data in the sample indicates the presence of defects. The observed weak anisotropy of the magnetic susceptibility at low temperatures is evidence of local distortions of the cubic crystal lattice of pyrochlore [10]. There was a violation of the selection rules in spectroscopic experiments performed on isostructural compounds [11].

The theory of random deformations caused by lattice point defects was built up in [12]. In the paper [13] the spectral properties of praseodymium zirconate were simulated based on the distribution function of random strains for an isotropic continuum [12] and a phenomenological model of electron-strain interaction. In the present paper, we simulated the spectral and thermodynamic properties of $Pr_2Zr_2O_7$, assuming the presence of lattice point defects that induce random deformations. In the calculations, we used the distribution function of random deformations, built up taking into account the elastic anisotropy of the crystal lattice [14].

**Results and Discussion**

In the presence of crystal lattice defects, the main electron doublet $\Gamma_3^+$ of the $Pr^{3+}$ ion in the $Pr_2Zr_2O_7$ lattice splits due to local symmetry reduction. The splitting value is determined by the local deformation, which is random in nature and depends on the concentration and type of defects, and the parameters of the electron-deformation interaction. The distribution function of random strains caused by point defects in the crystal lattice, taking into account elastic anisotropy, is the generalized Lorentz distribution for six independent components of the strain tensor of a cubic crystal [14]:

$$g(\boldsymbol{e}) = \frac{15\xi}{8\pi^3\gamma_{\mathrm{A}}\gamma_{\mathrm{E}}^2\gamma_{\mathrm{F}}^3}\left\{\gamma_A^{-2}e\left(\mathrm{A_{1g}}\right)^2 + \gamma_E^{-2}\sum_{\lambda=1}^{2}e_\lambda\left(\mathrm{E_g}\right)^2 + \gamma_F^{-2}\sum_{\lambda=1}^{3}e_\lambda\left(\mathrm{F_{2g}}\right)^2 + \xi^2\right\}^{-7/2}. \quad (1)$$

Here $e_\lambda(\Gamma)$ — linear combinations of the strain tensor components $e_{\alpha\beta}$ in the Cartesian coordinate system with the axes $X$, $Y$, $Z$ along the tetragonal axes of symmetry of the cubic lattice, transformed by the row $\lambda$ of the irreducible representation $\Gamma$:

$$\begin{aligned}
e_1(\mathrm{A_{1g}}) &= (e_{XX} + e_{YY} + e_{ZZ})/\sqrt{6},\\
e_1(\mathrm{E_g}) &= (2e_{ZZ} - e_{XX} - e_{YY})/\sqrt{12},\\
e_2(\mathrm{E_g}) &= (e_{XX} - e_{YY})/2,\\
e_1(\mathrm{F_{2g}}) &= (2e_{XY} - e_{ZX} - e_{YZ})/\sqrt{6},\\
e_2(\mathrm{F_{2g}}) &= (e_{XZ} - e_{YZ})/\sqrt{2},\\
e_3(\mathrm{F_{2g}}) &= (e_{XY} + e_{XZ} + e_{YZ})/\sqrt{3}.
\end{aligned} \quad (2)$$

The values of the parameters $\gamma_{\mathrm{A}}$=0.92, $\gamma_{\mathrm{E}}$=33.01, $\gamma_{\mathrm{F}}$=34.47, which characterize the difference in the width of the distribution function for strains transformed according to the irreducible representations $A_{1g}$, $E_g$ and $F_{2g}$ of the cubic group, respectively, were calculated by us according to the method described in [14] using the elastic constants of the crystal $La_2Zr_2O_7$ [15].

The $\xi = |\Omega_0| C_{\mathrm{d}}/48\pi$ parameter characterizes the width of the distribution function, which is proportional to the concentration of $C_{\mathrm{d}}$ defects and "to the strength" of defects (the ratio of the change in the lattice cell volume ν to the number of defects) $\Omega_0 = v^{-1}dv/dC_{\mathrm{d}}$, $v = a^3/4$, $a$ – cell parameter. The $\xi$ parameter was overviewed as a variable adjustable parameter.

The observed physical quantities $\langle A(T)\rangle$ were calculated as follows: the quantum-mechanical and quantum-statistical averages $A(T,\boldsymbol{e})$ were calculated for fixed values of the components of the strain tensor $\boldsymbol{e}$ and temperature $T$ and then averaged with the distribution function of random strains:

$$\langle A(T)\rangle = \int A(T,\boldsymbol{e})g(\boldsymbol{e})d\boldsymbol{e}. \quad (3)$$

The following Hamiltonian of the $Pr^{3+}$ ion was used in the calculations:

$$H = H_0 + H_{\mathrm{CF}} + H_Z + H_{\text{el-def}}, \quad (4)$$

where $H_0$ — Hamiltonian of a free ion, $H_{\mathrm{CF}}$ — Hamiltonian of an ion in a crystal field of a perfect crystal, $H_Z$ — energy of interaction of electrons with a local magnetic field, $H_{\text{el-def}}$ — Hamiltonian of interaction of 4f–electrons with lattice deformations.

The Hamiltonian $H_0$ of the free $Pr^{3+}$ ion was written in the space of 91 states of the $4f^2$ ground electronic configuration and included the energy of the electrostatic interaction between 4f electrons, the energy of the spin-orbit interaction and the spin-alien orbit interaction, the energy of the interconfiguration interaction [16].

The Hamiltonian $H_{\text{CF}}$ in the local coordinate system with the trigonal symmetry axis $z$ directed from the center of the tetrahedron formed by RE ions, to which the $Pr^{3+}$ ion under study belongs, to the corresponding vertex, is determined by six parameters $B_p^k$ of the CF:

$$H_{\text{CF}} = B_2^0 O_2^0 + B_4^0 O_4^0 + B_4^3 O_4^3 + B_6^0 O_6^0 + B_6^3 O_6^3 + B_6^6 O_6^6 , \quad (5)$$

where $O_p^k$ — linear combinations of spherical tensor operators similar to Stevens operators [17] in the basis of angular momentum eigenfunctions.

Several sets of CF parameters on the $Pr^{3+}$ ion in the $Pr_2Zr_2O_7$ [1, 18, 19] lattice have been proposed in the literature (Table 1). We did not use the previously proposed sets of CF parameters, since they were obtained either by considering the $H_{\text{CF}}$ operator in the truncated basis of the ground multiplet without taking into account the mixing of the wave functions of the ground and excited multiplets, or by fitting the calculated energy levels to the experimental least squares method and have physically unjustified values (in particular, signs that are opposite to the results of calculations), or they predict incorrect values of the effective magnetic moment and the $g$ factor. The calculations performed in the papers [1,18] predict the structure of the energy spectrum, which is inconsistent with the Raman scattering spectroscopy data [20].

We calculated the CF parameters on the $Pr^{3+}$ ion in the $Pr_2Zr_2O_7$ lattice as part of the semiphenomenological model of exchange charges [21]. The resulting set of parameters was then corrected by analyzing the spectra of inelastic neutron scattering [1] and the spectra of Raman light scattering [20]. The set of CF parameters for the $Pr^{3+}$ ion in the local coordinate system with the $z$ axis along the [111] trigonal symmetry axis of the cubic lattice and the $x$ axis along the $\mathbf{z}\times(-\mathbf{X}+\mathbf{Y})$ vector used in further calculations is given in Table 1.

The energies of sublevels of the ground multiplet calculated by us are compared with the experimental data in Table 2. Table 2 lists the irreducible representations of the point symmetry group $D_{3d}$, according to which the corresponding wave functions are transformed. The quasidoublet with frequencies 442 and 460 $cm^{-1}$ [20] observed in the Raman spectrum is due to the quasi-resonant interaction of the electronic excitation of $Pr^{3+}$ ions from the ground state to the first excited doublet ($\Gamma_3^+ \to \Gamma_3^+$) with the active doublet in the Raman scattering spectrum optical phonon of symmetry $F_{2g}$ with frequency 450 $cm^{-1}$ [15]. The calculated values of the effective magnetic moment $\mu_{\text{eff}}$ = 2.49 $\mu_B$ and the longitudinal component g of the ground state tensor (doublet $\Gamma_3^+$) $g_{\|}$ = 4.75 are in agreement with the values $\mu_{\text{eff}}$ = 2.45 $\mu_B$ [4] and $g_{\|}$ = 4.78 [23] determined from the experiments.

The resulting values of the semiphenomenological parameters of exchange charges model were used to calculate the parameters of the electron-deformation interaction (EDI). The Hamiltonian of the EDI of a RE ion has the form

$$H_{\text{el-def}} = \sum_{\Gamma,\lambda} V_\lambda(\Gamma) e_\lambda(\Gamma), \quad (6)$$

where $V_\lambda(\Gamma) = B_{p,\lambda}^k(\Gamma) O_p^k$ — electronic operators. The procedure for calculating the EDI parameters is described in [25]. The set of 30 independent $B_{p,\lambda}^k(\Gamma)$ parameters for $Pr_2Zr_2O_7$ slightly differs from the corresponding set for terbium ions in the $Tb_2Ti_2O_7$ crystal, which is given in [25].

As part of the effective spin Hamiltonian ($S$ = 1/2), the contribution of the EDI is described by the Hamiltonian $H_{\text{el-def}} = V\sigma_+ + h.c.$, $\sigma_+ = (\sigma_x + \sigma_y)/2$, $\sigma_\alpha$ are Pauli matrices. The linear function of the components of the strain tensor

$$V = V(\mathrm{E}_g)\left(e_1(\mathrm{E}_g) - ie_2(\mathrm{E}_g)\right) + V(\mathrm{F}_{2g})\left(e_1(\mathrm{F}_{2g}) + ie_2(\mathrm{F}_{2g})\right) \tag{7}$$

is determined by two parameters (the matrix elements of the corresponding operators $V_\lambda(\Gamma)$ on the wave functions of the ground doublet), the values $V(\mathrm{E_g})$ = 242 cm$^{-1}$ and $V(\mathrm{F_{2g}})$ = 1386 cm$^{-1}$ of which were obtained as part of the microscopic CF model with six parameters given in Table 1. The value of splitting of the ground doublet due to deformations is equal to

$$\Delta(\boldsymbol{e}) = 2\left\{\left[-V(\mathrm{E}_g)e_1(\mathrm{E}_g) + V(\mathrm{F}_{2g})e_1(\mathrm{F}_{2g})\right]^2 + \left[V(\mathrm{E}_g)e_2(\mathrm{E}_g) + V(\mathrm{F}_{2g})e_2(\mathrm{F}_{2g})\right]^2\right\}^{1/2}. \tag{8}$$

Praseodymium has a stable isotope $^{141}$Pr with nuclear spin $I$ = 5/2 (natural abundance 100%). When calculating the nuclear contribution to the heat capacity, the Hamiltonian (4) of the $Pr^{3+}$ ion also included the hyperfine interaction operator $H_{\text{HF}}$, the explicit form of which is given in [12].

We have obtained an assessment for the width of the random strain distribution function based on an analysis of experimental data on inelastic neutron scattering. The position of the line observed in the low-energy part of the inelastic neutron scattering spectrum in the absence of an external magnetic field depends on the sample and lies in area 2–3 cm$^{-1}$ [4, 7, 13]. The shape of the scattering intensity line at low temperatures can be described as the sum of the elastic scattering intensity, approximated by the Gaussian form function $I_{\text{elast}}(E) \sim \exp(-E^2/2\delta^2)$ with standard deviation $\delta$ = 0.376 cm$^{-1}$, and the inelastic scattering intensity $I_{\text{inelast}}(E)$, averaged over the splittings $\Delta(\boldsymbol{e})$ of the ground doublet induced by random deformations:

$$I_{\text{inelast}}(E) \sim \int \exp\left[-(E - \Delta(\boldsymbol{e}))^2/2\delta^2\right] g(\boldsymbol{e})d\boldsymbol{e}\,. \tag{9}$$

Fig. 1 shows the results of modeling the spectrum envelope of inelastic neutron scattering at a temperature of 1.4 K taking into account random deformations. It can be seen that the calculations reproduce well the observed neutron scattering intensity profile at the value of the variable $\xi$ = 2 · 10$^{-5}$ parameter. This value gives an upper bound for the $\xi$ parameter, since we neglected the magnetic and quadrupole interactions between the $Pr^{3+}$ ions. The peak in the intensity profile (line 3 in Fig. 1) arises as a consequence of the specific structure of the multidimensional strain distribution function.

Comparison of the experimental dependences of the heat capacity of a crystal $Pr_2Zr_2O_7$ [1–4,9] proves the influence of crystal lattice defects on its thermodynamic properties. Fig. 2 and 3 show the results of modeling the temperature dependences of the heat capacity. The calculations, the results of which are shown in Figs 2 and 3, were carried out with the parameter $\xi$ = 2 · 10$^{-5}$ determined from an analysis of the inelastic neutron scattering spectra. The heat capacity of the crystal $Pr_2Zr_2O_7$ was calculated as the sum of the magnetic and lattice contributions, the latter being assumed equal to the heat capacity of the nonmagnetic isostructural compound $La_2Zr_2O_7$ [2]. The magnetic contribution to the heat capacity was calculated by the formula

$$C_{p,\text{mag}} = \frac{N_{\text{A}}}{k_{\text{B}}T^2}\left\langle \mathrm{Tr}[H_s^2\rho] - \left(\mathrm{Tr}[H_s\rho]\right)^2 \right\rangle, \tag{10}$$

where ρ — ion density matrix $Pr^{3+}$ with Hamilton operator $H_s = H_0 + H_{\text{CF}} + H_{\text{el-def}} + H_{\text{HF}}$, $N_{\text{A}}$ — Avogadro number, $k_{\text{B}}$ — Boltzmann constant. It can be seen that the application of the theory of random deformations allows to qualitatively describe the Schottky anomaly observed near 2 K.

Fig. 3 shows the heat capacity $Pr_2Zr_2O_7$ in the low-temperature area (T ~ 0.1 K). Heat capacity calculations for a perfect crystal (line 1 in Fig. 3) predict a Schottky anomaly in this area, due to the hyperfine structure of the main non-Kramers doublet, with a maximum at a temperature much higher than the temperature of this maximum according to the measurement data (symbols in Fig. 3). The interaction of ions with random deformations suppresses the hyperfine interaction and significantly shifts the corresponding Schottky anomaly to the area of ultralow temperatures (line 2 in Fig. 3).

Thus, the use of the theory of random deformations allowed to qualitatively reproduce the low-energy profile of the inelastic neutron scattering spectrum and features in the temperature dependences of the heat capacity of praseodymium zirconate at low temperatures.

In order to estimate the concentration of point defects, which corresponds to the obtained value of the width of the strain distribution function, one can use the "stuffed pyrochlore" ($Pr_{2+x}Zr_{2-x}O_{7-x/2}$) model and X-ray diffraction data. As part of this model, we get $|\Omega_0| C_d = (3|x|/a) da/dx$, where $a$ = 10.677 Å [10], $da/dx$ = 0.37 Å [9], then $\xi \approx 7 \cdot 10^{-4} \cdot |x|$. Substituting the value of the width of the distribution function $\xi = 2 \cdot 10^{-5}$, which was determined from a comparison of the results of calculations and experimental data, we obtain the assessment $|x| \approx 0.03$, which corresponds to a defect concentration of approximately 1.5%. It is noted in the literature that for RE pyrochlores, samples with a 2% deviation from stoichiometricity are considered to be of sufficient quality. Thus, the application of the theory of random deformations allowed to indirectly assess the quality of the samples by determining the width of the deformation distribution function from the simulation of the available experimental data.

## Conclusions

Our calculations show the importance of taking into account random deformations caused by defects in the crystal lattice when studying the properties of pyrochlores with non-Kramers ions at low temperatures. The effects caused by the influence of displacements of oxygen sublattices corresponding to lattice vibrations of symmetry $F_{2g}$, active in the spectrum of Raman light scattering, on the coupling parameters of the electron subsystem with macroscopic deformations require additional study. In addition, there may be additional significant effects due to multipole interactions between RE ions through the phonon field and two-particle terms in the electron-deformation interaction. From the point of view of experimental research, it is of interest to measure the speed of sound and forced magnetostriction at low temperatures and in weak magnetic fields for pyrochlores containing $Pr^{3+}$ and $Tb^{3+}$ non-Kramers ions. The analysis of the data from these measurements will allow to correct the set of electron deformation interaction parameters for further study of the defects effect on the properties of pyrochlores containing non-Kramers RE ions.

## Acknowledgments

The authors are grateful to B.Z. Malkin for scientific guidance, discussion of the results obtained, and editing of the manuscript.

This study was supported by grant No 12-00244 from the Russian Science Foundation.

**Table 1.** Parameters of the CF on the $Pr^{3+}$ ion in the $Pr_2Zr_2O_7$ crystal ($cm^{-1}$). For comparison, the CF parameters on the $Pr^{3+}$ on in isostructural compounds $Pr_2Hf_2O_7$ and $Pr_2Sn_2O_7$.

| | $B_2^0$ | $B_4^0$ | $B_4^3$ | $B_6^0$ | $B_6^3$ | $B_6^6$ | Reference |
|---|---|---|---|---|---|---|---|
| $Pr_2Zr_2O_7$ | 355 | 397 | −3233 | 87.5 | −301 | 574 | [1] |
| | 255 | 466 | −3979 | 143 | 1359 | 1197 | [18] |
| | 193 | 369 | −5336 | 62.3 | 475 | −429 | [19] |
| | 252 | 355 | −5122 | 32.3 | −193 | −252 | [19] |
| | **157** | **428** | **−4335** | **79.4** | **1140** | **1235** | This work |
| $Pr_2Sn_2O_7$ | 233 | 436 | −3835 | 72.7 | 1109 | 1471 | [22] |
| $Pr_2Hf_2O_7$ | 136 | 429 | −4592 | 83.2 | 1154 | 1367 | [23] |
| | 138 | 406 | −4070 | 74.1 | 1102 | 1189 | [24] |

**Table 2.** Calculated and measured energy levels of the ground multiplet $^3H_4$ of the $Pr^{3+}$ ion ($cm^{-1}$).

| Calculation | | Experiment | | | |
|---|---|---|---|---|---|
| $Pr_2Zr_2O_7$ | | | | $Pr_2Sn_2O_7$ | $Pr_2Hf_2O_7$ |
| This work | | [1] | [20] | [22] | [24] |
| $\Gamma_3^+$ | 0 | 0 | 0 | 0 | 0 |
| $\Gamma_1^+$ | 81.0 | 76.5 | 76.6 | 144 | 74.2 |
| $\Gamma_3^+$ | 453 | 460 | 442, 460 | 466 | 431 |
| $\Gamma_1^+$ | 671 | 660 | 662 | 663 | 634 |
| $\Gamma_3^+$ | 758 | 751 | 761 | 806 | 721 |
| $\Gamma_2^+$ | 875 | 878 | 879 | 927 | 836 |

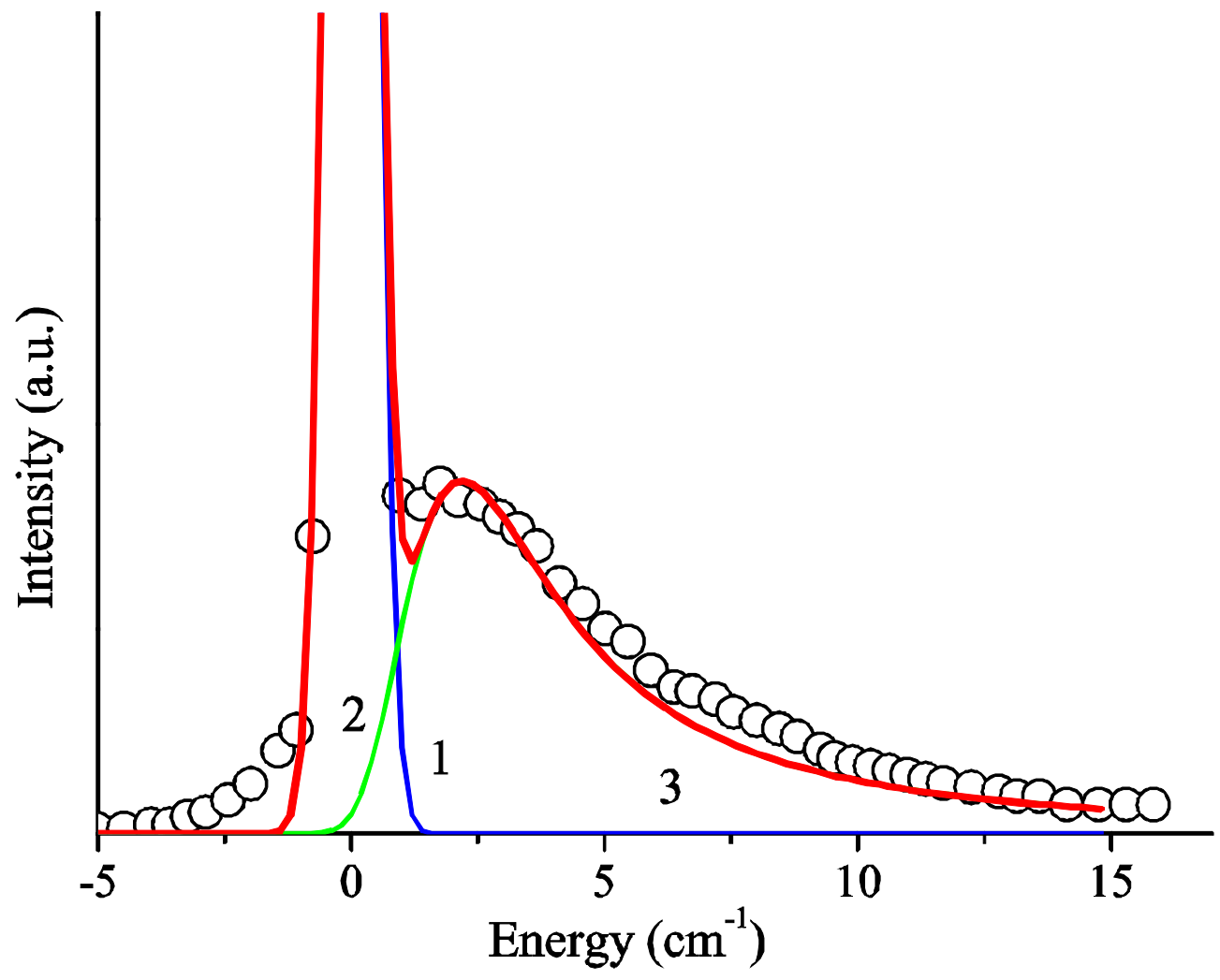

**Figure 1.** The measured (symbols) and calculated (3 line) neutron scattering intensity as a function of the transfer energy, $T$ = 1.4 K. The lines 1 and 2 show the elastic and inelastic contributions, respectively. The experimental points were digitized according to the measurement data in [7].

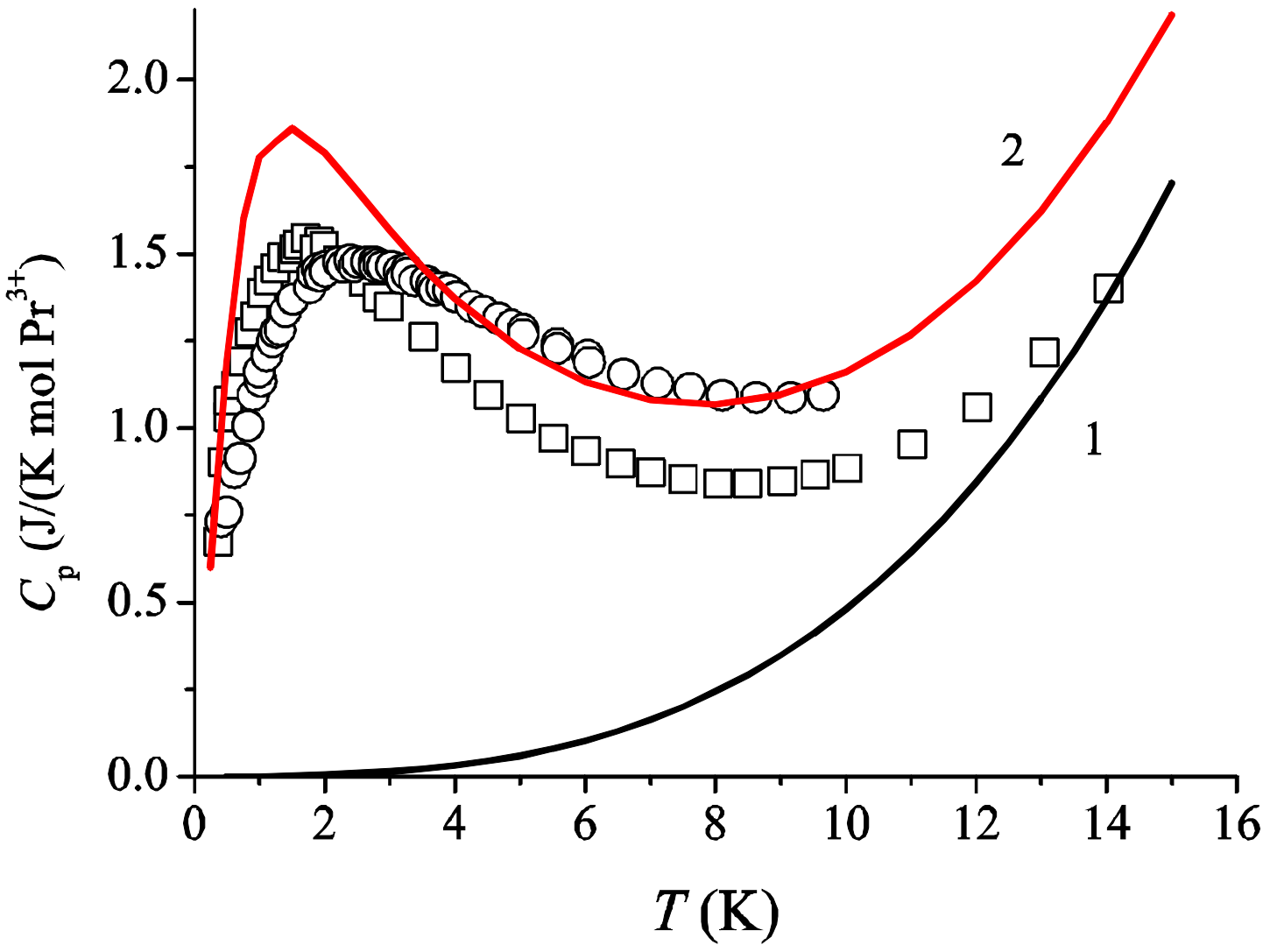

**Figure 2.** The measured (symbols) and calculated (curve 1 — perfect crystal lattice, 2 — lattice with random defects) heat capacity as a function of temperature. Experimental data taken from [2] (squares) and [4] (circles).

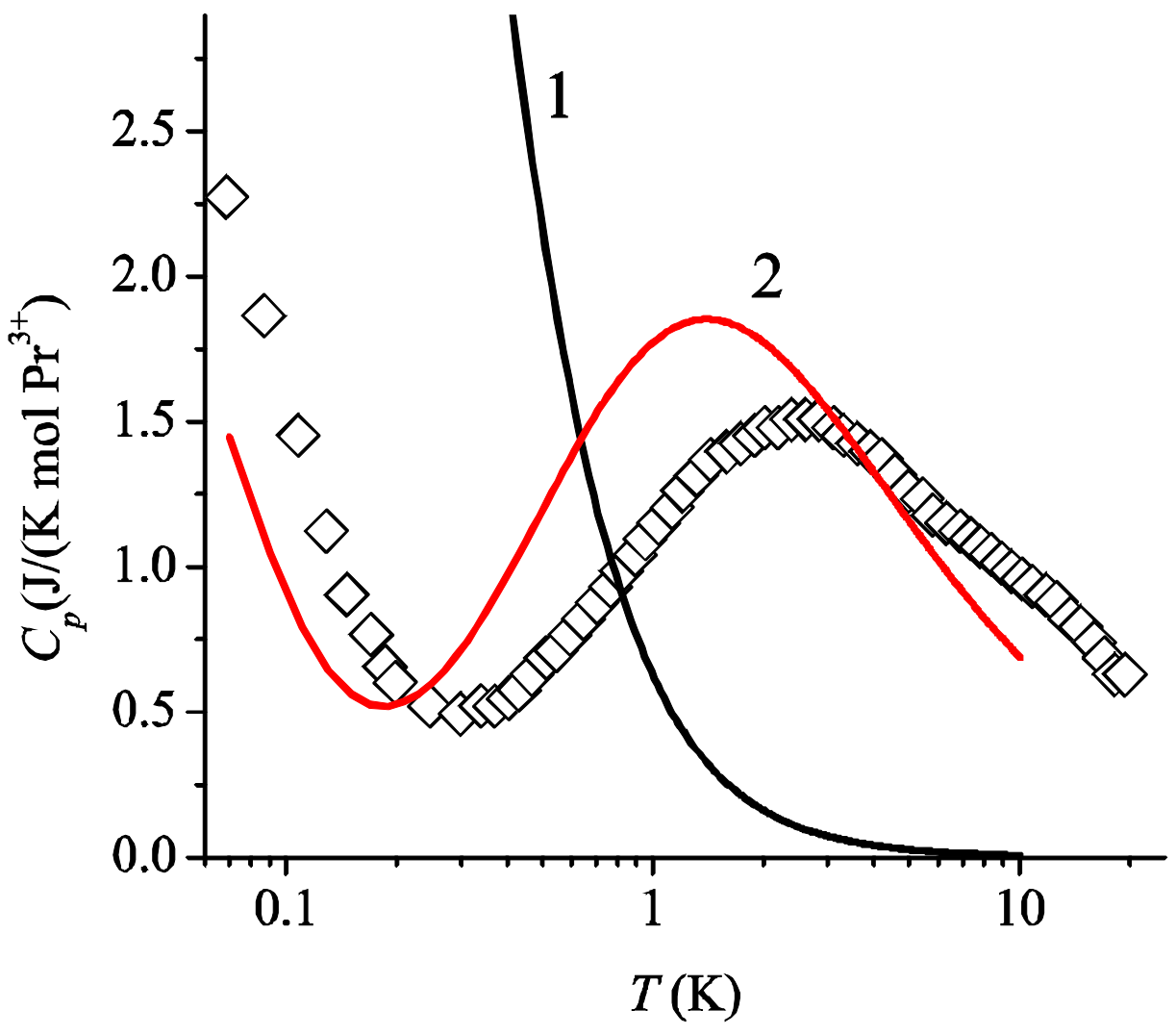

**Figure 3.** The measured (symbols) temperature dependence of the heat capacity and the calculated sum of the magnetic and nuclear contributions to the heat capacity (curve 1 — perfect crystal lattice, 2 — lattice with defects). The experimental data was taken from [1].